\documentclass{statsoc}

\usepackage[a4paper]{geometry}
\usepackage{graphicx}
\usepackage[textwidth=8em,textsize=small]{todonotes}
\usepackage{amsmath}
\usepackage{natbib}
\usepackage{tcolorbox}
\usepackage{float}
\usepackage{url}

\usepackage{etoolbox}
\usepackage{enumerate}

\title[Projecting COVID-19]{``Back to the future'' projections for COVID-19 surges}
\author[J. Sunil Rao {\it et al.}]{J. Sunil Rao}
\address{Division of Biostatistics, University of Miami,
Miami,
USA.}
\email{jrao@miami.edu}
\author{Tianhao Liu}
\address{Division of Biostatistics, University of Miami,
Miami,
USA.}
\author{Daniel Andr\'es D\'iaz-Pach\'on}
\address{Division of Biostatistics, University of Miami,
Miami,
USA.}

\begin{document}
\begin{abstract}
We argue that information from countries who had earlier COVID-19 surges can be used to inform another country's current model, then generating what we call {\it back-to-the-future} (BTF) projections.  We show that these projections can be used to accurately predict future COVID-19 surges {\it prior to an inflection point of the daily infection curve}.   We show, across 12 different countries from all populated continents around the world, that our method can often predict future surges in scenarios where the traditional approaches would always predict no future surges.  However, as expected, BTF projections cannot accurately predict a surge due to the emergence of a new variant.  To generate BTF projections, we make use of a matching scheme for asynchronous time series combined with a response coaching SIR model.  
\end{abstract}

\section{Introduction}
``The past is prologue'' (Shakespeare). ``The best predictor of future behavior is past behavior'' (Twain). ``The best way to predict the future is to study the past or prognosticate'' (Kiyosaki).  These are all famous quotes which, when applied to important prediction or projection problems (projection being prediction into the future), suggest that a careful understanding of past events is essential to predicting future trends. 

Yet, when applied to the problem of projecting a new surge of COVID-19 infections in India, back in mid February 2021, these strategies did not work.  India had seen a remarkable downturn in their daily new cases curve and all models built at that time were projecting a continuing trend in that direction, down towards zero daily new cases.  All sorts of explanations were produced to why India escaped relatively unscathed, including cross-protection from other regular vaccines, like the BCG TB vaccine; a younger age distribution to the population; a warmer climate; and more homes with open window settings \citep{Mallapaty21}.

But by late March or early April, a significant upturn in the daily new cases curve had taken hold and India was rapidly experiencing a second surge dramatically more ferocious than the first.  In fact, daily new cases counts would cross the 400K per day soon thereafter (reported cases granted and likely hugely under-counted), with a lagging rise in the number of hospitalizations and deaths.  

So if modeling using the first surge data was not informative, was there any way to objectively predict the second surge? And to make things even more challenging, can a future surge be predicted before that surge has actually started?  That is, prior to the inflection point between the ending of a current surge and the start of a new one.  We argue surprisingly that there may be.  In this paper we will present a method we call {\it back-to-the-future} (BTF) projections that borrows information from so-called ``matching'' countries who experienced an earlier surge.  This information is used to coach projections forward in time.   

This paper is organized as follows.  We begin with a short review of the basic modeling strategies for pandemic data and why projections are so sensitive to the point of inflection.  We then introduce the BTF idea and algorithm for fitting.  Empirical results on 12 different countries from every continent except Antarctica are presented with comparisons against the basic modeling approaches.  We finally provide some justification for the matching and coaching used in making BTF projections.

\section{Contrasting modeling strategies for pandemics}
\subsection{Compartment Models}
The SIR model is the simplest compartment model for describing the evolving dynamics of an epidemic through a population.  It can be described by a set of ordinary differential equations (ODEs), 

\begin{align}\label{DiffEqs}
\begin{aligned}
	\frac{dS}{dt} &=   - \frac{\beta I(t)S(t)}{N}, \\
	\frac{dI}{dt} &=  \frac{\beta I(t)S(t)}{N} - \gamma I(t), \\
	\frac{dR}{dt} &=  \gamma I(t),
\end{aligned}
\end{align}
where, at time $t$, $S$ is the susceptible population, $I$ is the number of infectious, $R$ is the number removed either by death or recovery, and $N$ is the sum of these three:
\begin{align*}
	S(t) + I(t) + R(t) = N.
\end{align*}

The parameters $\beta$ and $\gamma$ are the transmission and recovery rates, respectively.  From \eqref{DiffEqs},
\begin{align*}
	\frac{dS}{dt} + \frac{dI}{dt} + \frac{dR}{dt} = 0.
\end{align*}

Also from \eqref{DiffEqs}, dividing the first equation by the third, and integrating with respect to $S$ and $R$, 
\begin{align*}
	S(t) = S(0)e^{-R_{0}(R(t) - R(0))/N},
\end{align*}
where $R_{0}$ is the basic reproduction number given by $R_{0} = \beta/\gamma$.  At the outset of an epidemic, when $S \approx N$, infection numbers begin to surge as $R_{0} \gg 1$.  Subsequent surges are characterized by the ratio $N/S$.  When $R_{0} > N/S$, infection numbers rise more rapidly, hit a peak when $R_{0} = N/S$, and then decline as $R_{0} < N/S$.   

When assumed purely mechanistic, numerical methods such as Euler discretization or the Runge-Kutta approximation method \citep{Butcher16} can be used to obtain approximate solutions of the ODEs with given boundary conditions.  In a statistical analysis framework, a model is constructed with a deterministic and random component.  The former is the SIR model itself.  The latter allows for a random sampling scheme, thus creating a stochastic extension of the mechanistic model.  Parameter estimation can be done via frequentist optimization, like least squares, the method of moments, maximum likelihood estimation, or Bayesian approaches using Markov Chain Monte Carlo techniques.  A clear advantage of the stochastic extensions is the ability to quantify uncertainty in parameter estimation and prediction due to sampling variability.  A full account of the SIR model (and related compartment model extensions) can be found in \citet{Tang20}.  




\subsection{Time series ARIMA models}
Time series models have also been exploited for modeling epidemic data trends \citep{Alabdulrazzaq21, Song16}.  Using new notation, we will let the daily infection counts be $Y_t$ and  defined $\Delta^{d}Y_t = (1-L)^{d}Y_{t} = Y_{t}- Y_{t-d}$, where $d$ is the number of differences needed to make the series stationary, then the ARIMA($p,d,q$) model (Box and Jenkins 1976) has the form
\begin{align}\label{ARIMA}
	A(L)(1-L)^{d}Y_{t} = \delta + \Omega(L)\epsilon_{t},
\end{align}
where $A(L)=1-\alpha_{1}L - \ldots - \alpha_{p}L^{p}$, $\Omega(L)=1-\theta_{1}L-\ldots-\theta_{q}L^{q}$, $p$ is the autoregressive order, $q$ the moving average order, and $L$ is known as the backshift operator.  The random variable $\epsilon_{t}$ is white noise assumed to follow a normal distribution.  The $\alpha_1, \ldots, \alpha_p$ are the autoregressive parameters, and the $\theta_1, \ldots, \theta_q$ are the moving average parameters, both sets to be estimated by maximum likelihood.  The order of the ARIMA($p,d,q$) model is typically chosen using a model selection criterion, like BIC \citep{BIC78} or AIC \citep{AIC74}, among other methods.

\subsection{Curve fitting}
Curve fitting essentially amounts to deriving a functional relationship between $Y_t$ and $t$ such that the estimated curve matches the observed daily infection count trend as closely as possible. This approach is generally considered less tied to underlying assumptions about features within the population that might be driving the daily infection numbers.  However, the drawback is that it's not a mechanistic approach and thus may not do as well with longer term forecasts.  Some examples include the generalized logistic model \citep{Aviv-Sharon20} and the generalized Gaussian cdf \citep{Ciufolini20}, both adopted by the Institute of Heath Metrics and Evaluation (IHME).   These models have been extended to allow for incorporation of covariates that can connect different locations together (\url{https://ihmeuw-msca.github.io/CurveFit/methods/}).

\subsection{Surge prediction and the sensitivity to point of inflection}
The focus in this paper is to predict future COVID-19 surges {\it before} an inflection point for the surge itself --- in other words, on the downward trajectory of the previous surge or in a valley before the future surge.   

\begin{figure}[h]
\centering 
\includegraphics[width=.65\textwidth, height=7cm]{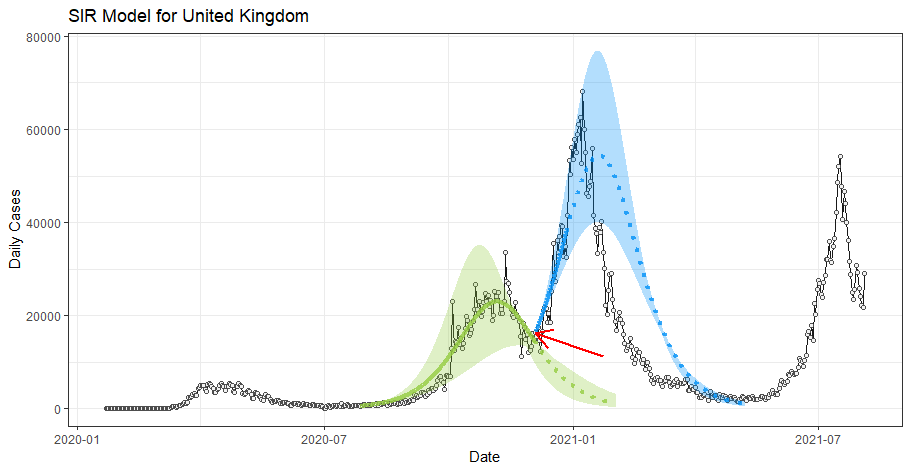}
\caption{Sensitivity of the UK SIR model projections to the inflection point (red arrow) of daily infections curve.}
\label{inflection}
\end{figure}

Projections before and after an inflection point can be markedly different.  To illustrate this point, consider SIR models fit to daily case count data for the United Kingdom, as shown in Figure~\ref{inflection}.  The red arrow on the plot indicates a point of inflection before the start of the third surge around November, 2021.  Let's suppose this is the surge we are trying to predict. The green curve is an SIR model fit to data prior to the inflection point that is been projected forward past the inflection point in Figure~\ref{inflection}.  Notice how it's descending to zero.  Now assume we wait some days to make the projection for the third surge.  The projected curve from such an SIR model would look like the blue curve in the figure.  As expected, it is rising upwards towards the observed peak daily count.  This looks to be much more accurate.  These types of projections are of less public health planning value, since with highly contagious viruses like the omicron variant of COVID-19, with an $R_0$ number estimated to be near 10, it is nearly impossible to blunt the surge after the point of inflection because one is always ``running behind'' the virus.  Our proposed methodology seeks to do better than the green curve based on the same observed data.

\section{Back to the future projections}
We now restrict our attention to one particular sequence --- the daily infection counts over time.  Our main interest is to project a future oncoming next surge during the downward trajectory of the current surge {\it but prior to an 
inflection point in the curve that might indicate the start of a new surge}.  As just shown, standard approaches will have all projected curves descending down towards zero daily counts.  

To improve naive projections, we exploit the very nature of a pandemic --- the fact that infections are spreading asynchronously in time across different countries.  Countries ($B_{m}, m=1,\ldots, M$) that have experienced a surge earlier in time may provide useful information in making projections forward in time for a country of interest ($A$).  This is done by estimating an ARIMA time series model for $A$ for the current surge  $S_{1}$ (say $\hat{f}_{t \in S_{1}}(A)$), shifting this curve backwards in time and overlaying its fitted curve with fitted curves from the other $B_{m}$ countries {\it previous} surges ($S_{1,m}$) (say $\hat{f}_{t \in S_{1,m}}(B_{m})$).  A determination of best country match is then made based upon the pairwise difference in fitted curves $\hat{f}_{t \in S_{1}}(A)$ and $\hat{f}_{t \in S_{1,m}}(B_{m})$.

Once the best match ($\tilde{B}_{m}$) amongst the $M$ countries has been established, an SIR model is fit to the {\it observed} daily infection counts forward in time for $\tilde{B}_{m}$. {\it This data is actually observed}, which is a key fact. Relevant SIR curve parameter estimates are then passed to country $A$, using $A$'s current initial conditions, to make {\it not yet observed} projections forward in time. This is a type of statistical {\it coaching}.  It's useful to impose a short ``washout'' period to allow the current surge to come to completion.  We call this {\it Back to the Future} (BTF) projections.  The steps can be summarized in the algorithm:

\begin{tcolorbox}
\begin{center}\underline{{\bf Back to the Future Projection Algorithm}}
\end{center}
\vskip10pt

Suppose we want to project country $A$ after $A$'s $i$-th surge.
\begin{enumerate}
  \item Select candidate countries such that
  \begin{enumerate}[i]
    \item - its $i$-th surge happens before $A$'s $i$-th surge,
    \item - it has sufficient data for $i$-th surge to match with $A$'s $i$-th surge (same length interval),
  \end{enumerate}
  Denote these candidate countries as $\{ B_k \}_{k = 1}^K$.
  \item Fit ARIMA models for the $i$-th surge of $A$ and $B_k$.  
  
  	Smooth these fits using cubic smoothing splines, with the degree of smoothing determined by leave-one-out cross-validation. 
  
  	Denote the fitted models by $\hat{A}$ and $\hat{B}_k$.
  \item From $\{ B_k \}_{k = 1}^K$, select the country most similar to $A$ by
  	\begin{align*}
  	\tilde{B}_m \equiv \text{arg\,min}_{B_k} \left\{ \text{median} \left[\hat{A}^{sd} - \hat{B}^{sd}_k\right] \right\},
  	\end{align*}
  	where $\hat{A}^{sd}, \hat{B}^{sd}_k$ be the standardization of $\hat{A}, \hat{B}_k$ by its maximum.
  
  \item Fit an SIR model to $\tilde{B}_m$ after its $i$-th surge (a 10-day gap may be introduced to washout the effect of the $i$-th surge).
  
  \item Pass the estimated parameters $\hat{\beta}(\tilde{B}_{m}), \hat{\gamma}(\tilde{B}_{m})$ by the SIR model of $\tilde{B}_m$ to the SIR model with $A$'s initial conditions. 
  
  \item Generate the projection using this new SIR model.
\end{enumerate}
\end{tcolorbox}

\subsection{Sensitivity analysis}

To analyze the sensitivity of the SIR model, we jitter the two parameters $\beta$ and $\gamma$ for a small amount $\delta$. In practice, we sample the parameter pair by a uniform distribution on $[\beta - \delta, \beta + \delta]
\times [\gamma - \delta, \gamma + \delta]$ (in our cases $\delta$ is chosen as 0.01). Then run the SIR model by each pair of these parameters. We  can shade the union of these individual runs.  

\subsection{Why coaching using a compartment model?}
The mechanistic nature of the systematic component of the compartment model provides a more parsimonious representation of country $\tilde{B}_{m}$ over a longer period of time. It also allows a clear path to incorporating country $A$'s specific characteristics. This helps to anchor the BTF projections and to generate more accurate projected trends over longer windows of time, rather than purely generating accurate short-term projections which are of limited public health benefit.  

Contrast this to coaching using an ARIMA model instead from country $\tilde{B}_{m}$. Since only lagged effects can be modeled in the ARIMA model, country $\tilde{B}_{m}$'s shape of their next surge (after the matching one), will not fully inform country $A$ forecasts of interest.  

\section{Data}
COVID-19 infection volume time series came from the Johns Hopkins University CSSE COVID-19  Tracking Project and Dashboard which when the data was pulled ranged from 01/22/2020 until 04/12/2021 (correct dates here).  

\section{Performance on a selection of countries}
A BTF analysis was carried out for a selection of 12 countries from all 6 populated continents around the world. Thus the performance of our methodology was examined regardless of the regional variation that might exist from continent to continent.   In particular, the chosen countries experienced second surges during our time window of analysis and the goal was to accurately forecast second surges from a lagged time point towards the end of their first surges (i.e. before the inflection point of the daily infection curve happened, indicating the start of a potential second surge).  This would be a truly honest projection and would more clearly demonstrate the utility of the BTF methodology. Usual forecasting with compartment models, ARIMA models, or curve fitting, would all indicate the projected curves continue downwards, given that the projections were made from a point in time on the downward trajectory of the first surges.  As a negative control, we also included Australia where no second surge was detected during the analysis time window. 

\begin{figure}[H]
\centering 
\includegraphics[width=.45\textwidth, height=5cm]{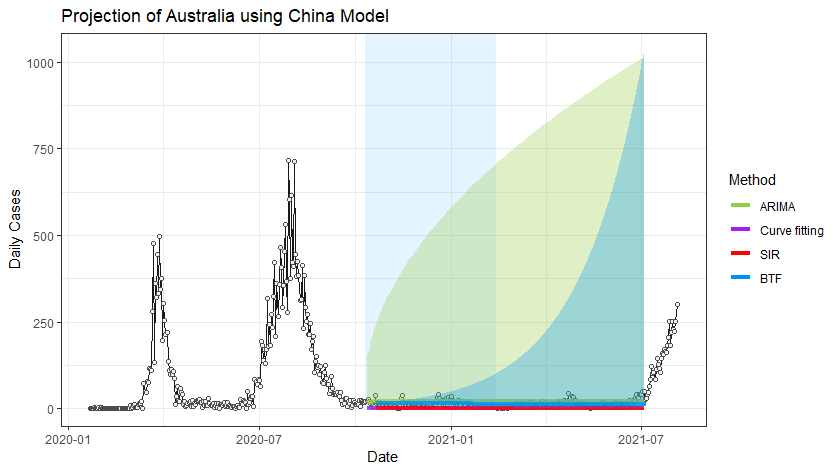}
\includegraphics[width=.45\textwidth,height=5cm]{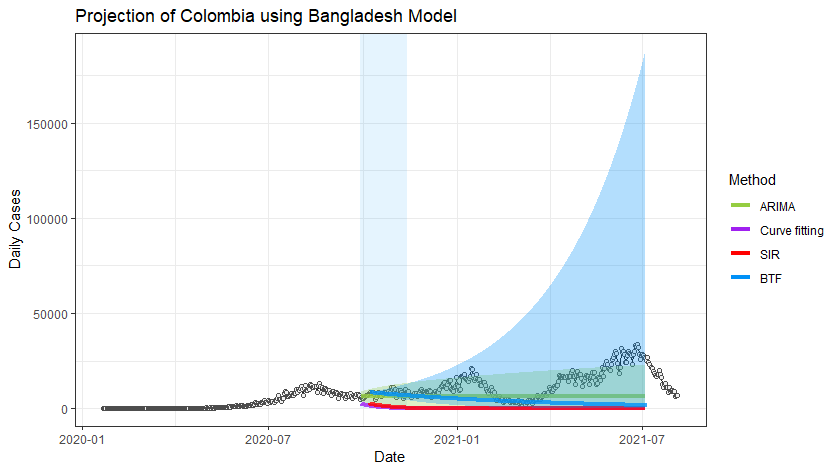}
\includegraphics[width=.25\textwidth, height=5cm]{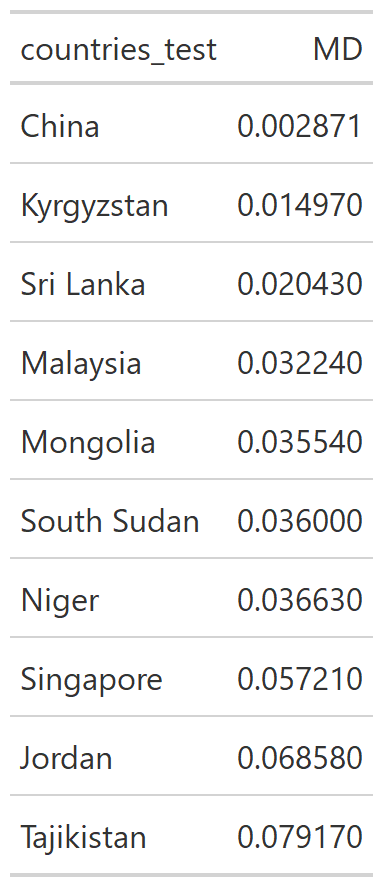} \hskip80pt
\includegraphics[width=.25\textwidth,height=5cm]{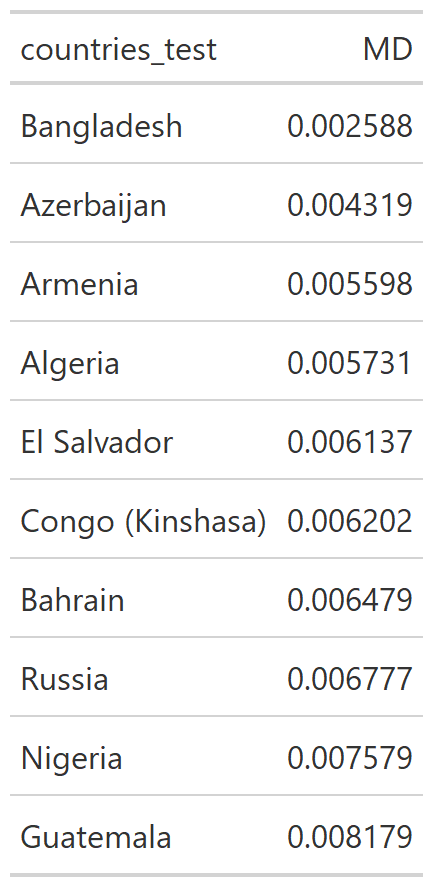}
\includegraphics[width=.45\textwidth,height=5cm]{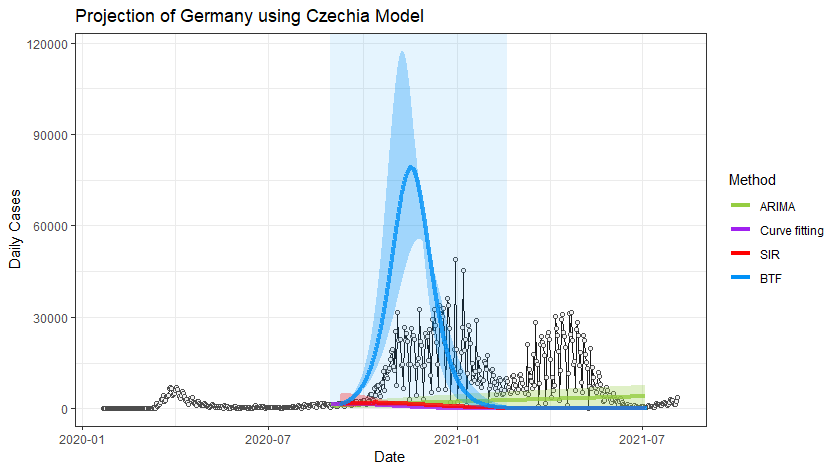}
\includegraphics[width=.45\textwidth,height=5cm]{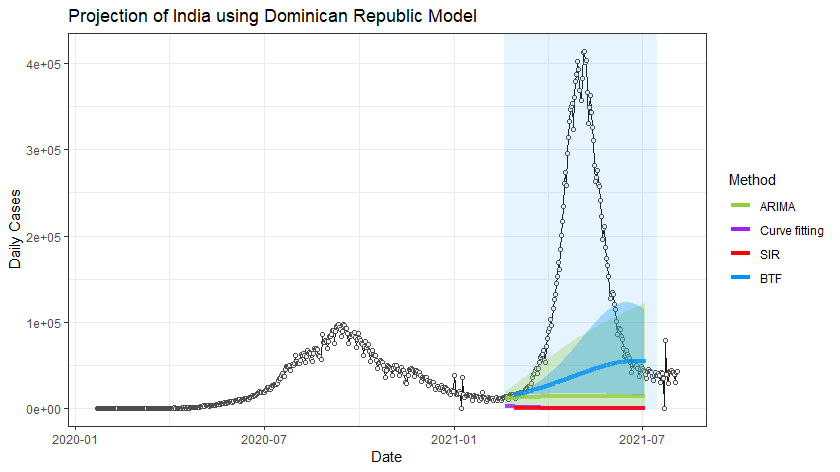}
\includegraphics[width=.25\textwidth,height=5cm]{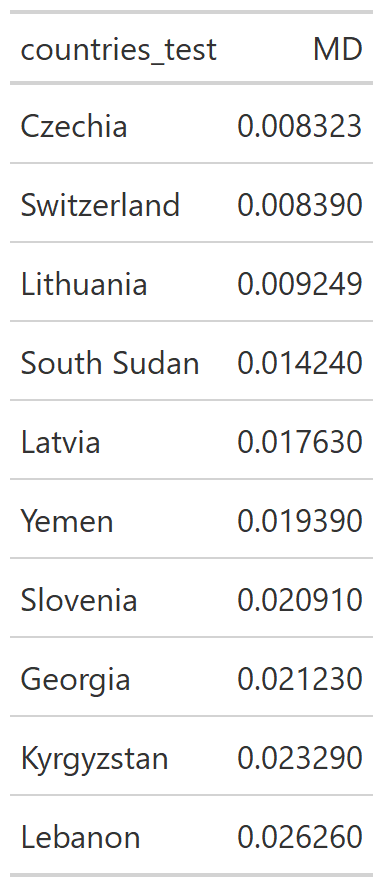}\hskip80pt
\includegraphics[width=.25\textwidth,height=5cm]{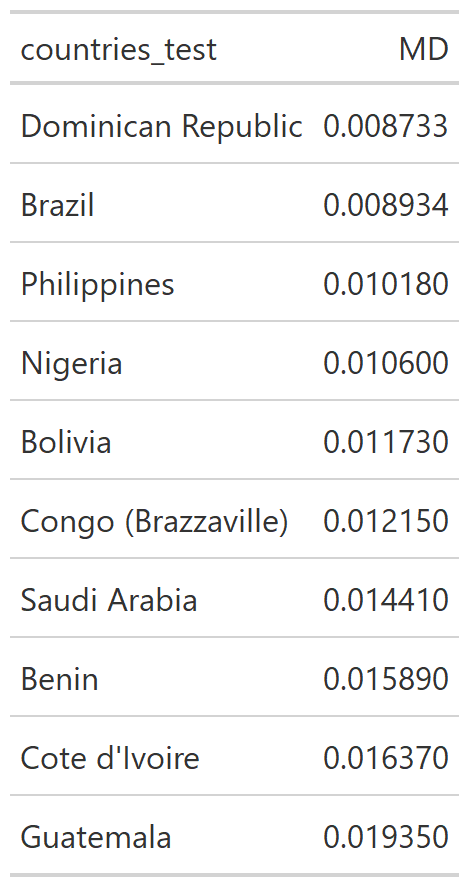}
\caption{Projection curves for Australia, Colombia, Germany, and India, using BFT projections.  Projected blue curve and region of projection (before inflection point) of next surge in shaded blue.  Note that a 10 day washout period is forced before projections start. Matching country ranking tables shown underneath each plot.  }
\label{group1}
\end{figure}

\begin{figure}[H]
\centering 
\includegraphics[width=.45\textwidth, height=5cm]{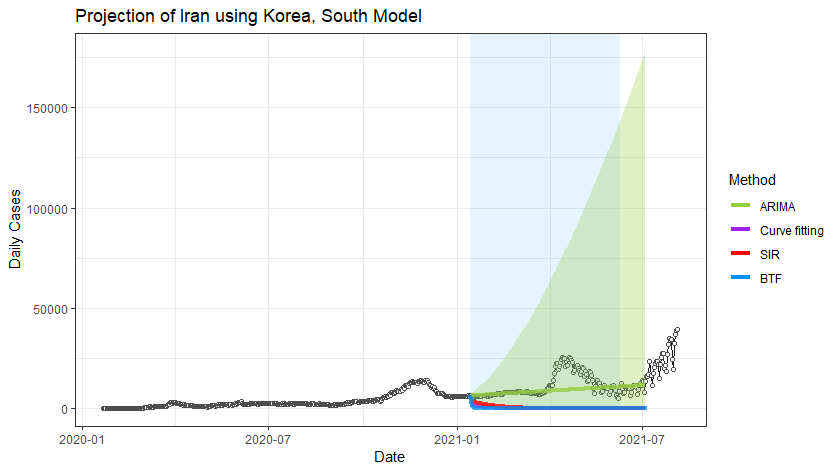}
\includegraphics[width=.45\textwidth,height=5cm]{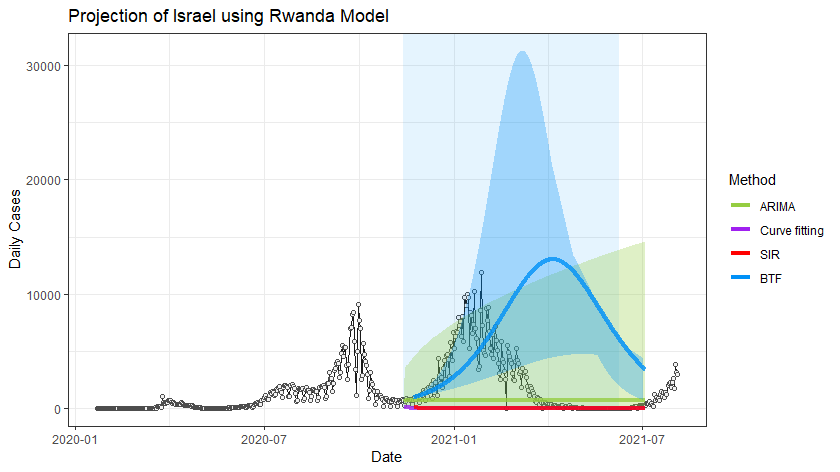}
\includegraphics[width=.25\textwidth, height=6cm]{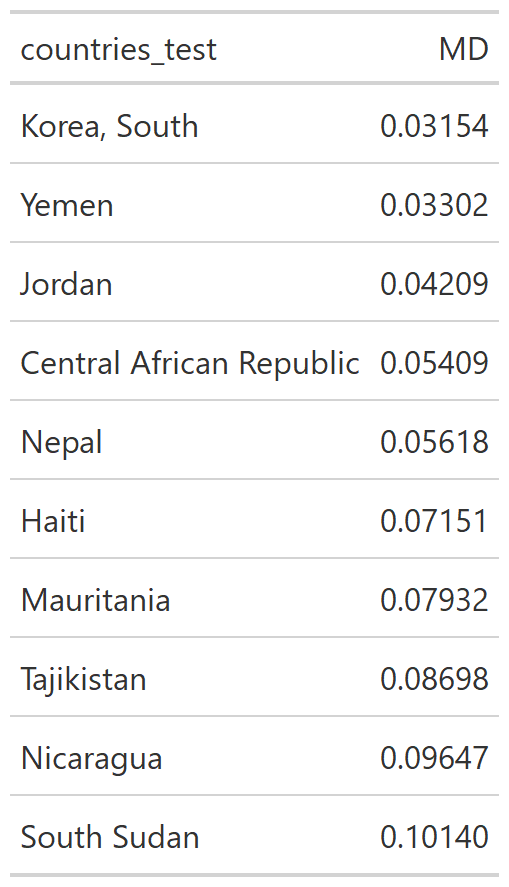}\hskip80pt
\includegraphics[width=.25\textwidth,height=6cm]{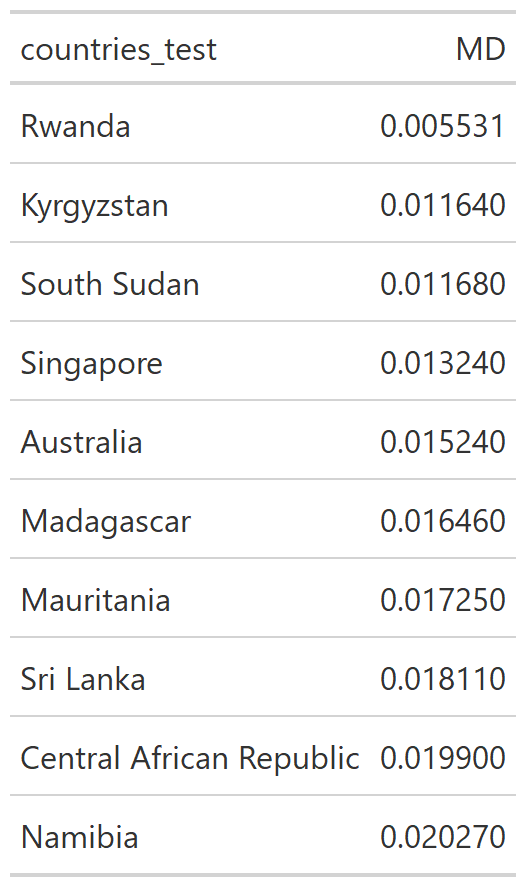}
\includegraphics[width=.45\textwidth,height=5cm]{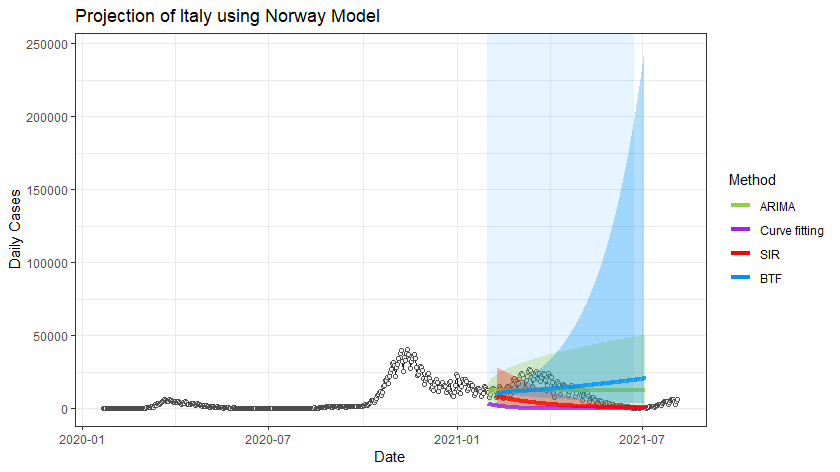}
\includegraphics[width=.45\textwidth,height=5cm]{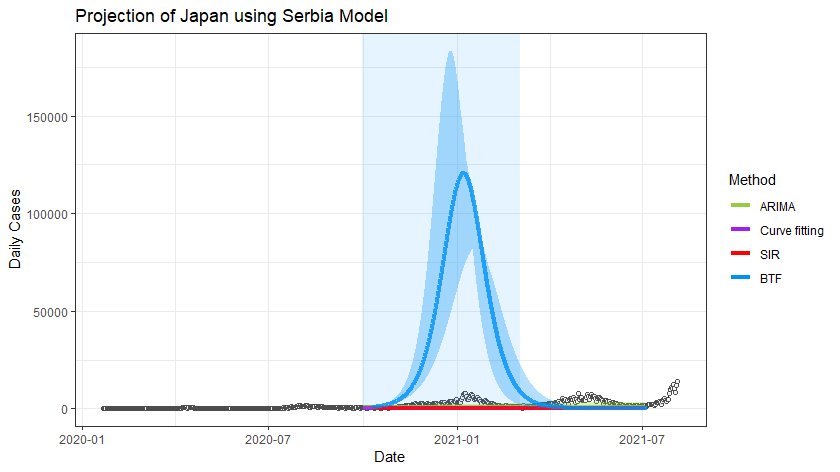}
\includegraphics[width=.25\textwidth,height=6cm]{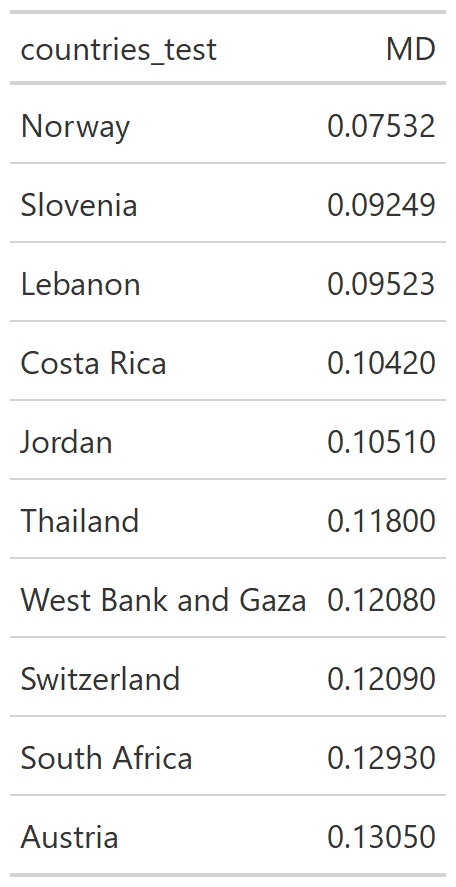}\hskip80pt
\includegraphics[width=.25\textwidth,height=6cm]{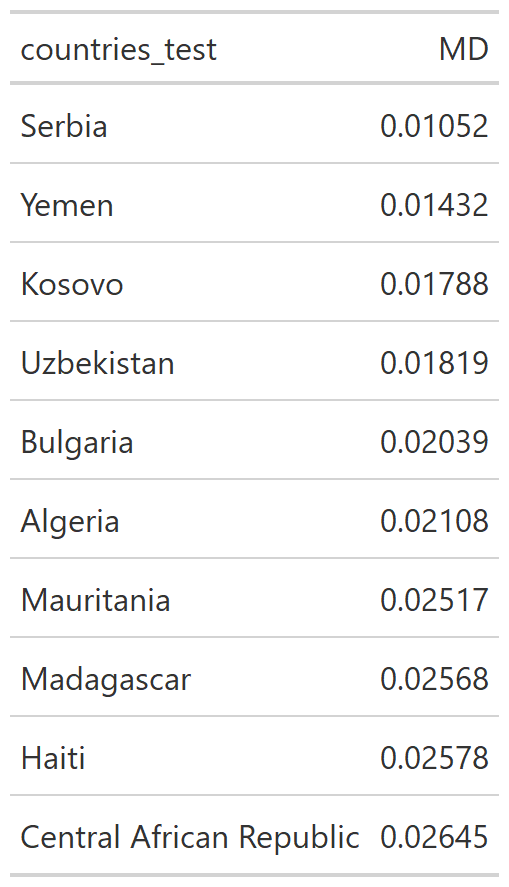}
\caption{Projection curves and matching country ranking tables for Iran, Israel, Italy, and Japan, using BFT projections.}
\label{group2}
\end{figure}

\begin{figure}[H]
\centering 
\includegraphics[width=.45\textwidth, height=5cm]{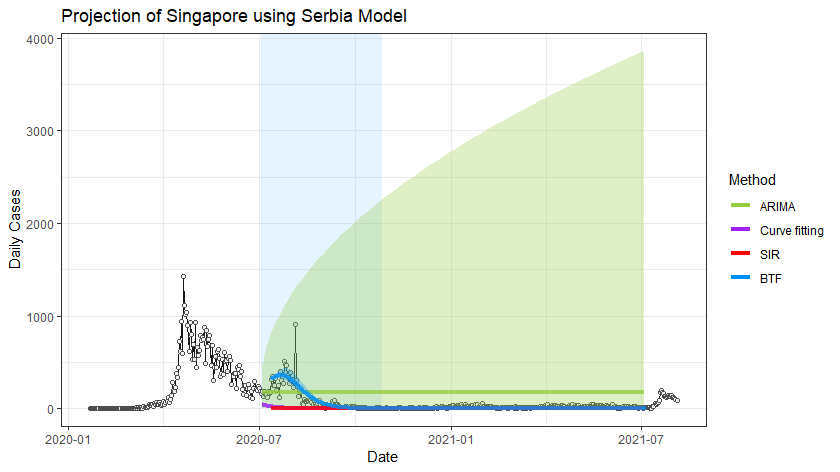}
\includegraphics[width=.45\textwidth,height=5cm]{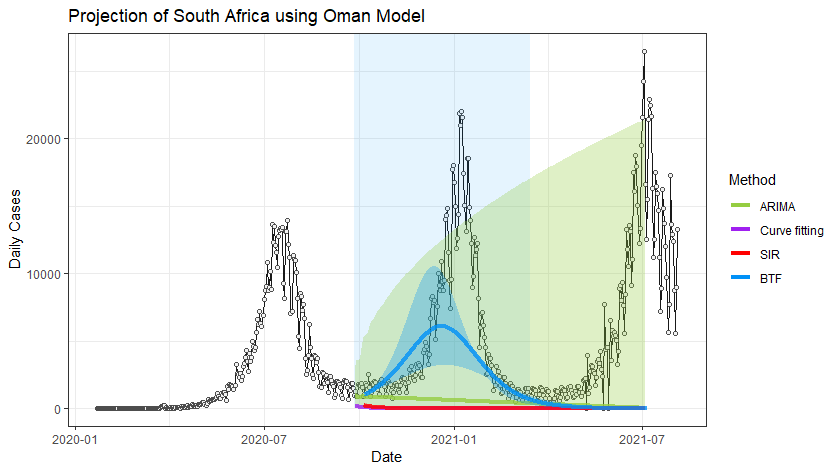}
\includegraphics[width=.25\textwidth, height=6cm]{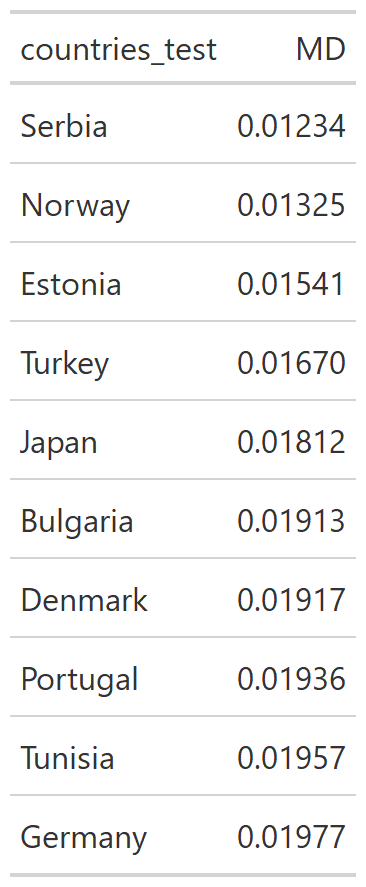}\hskip80pt
\includegraphics[width=.25\textwidth,height=6cm]{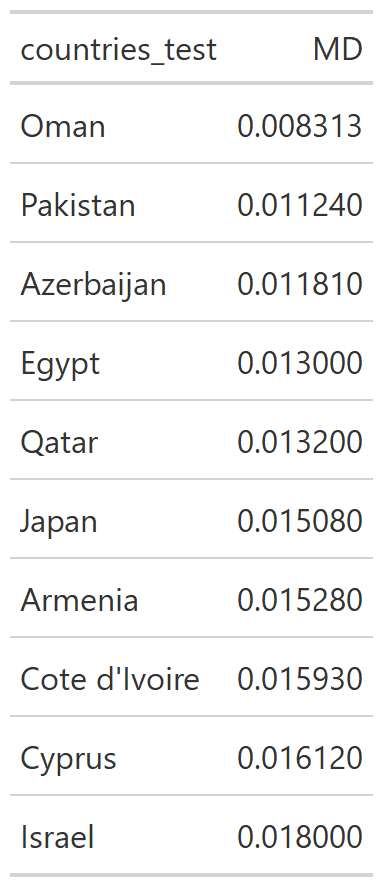}
\includegraphics[width=.45\textwidth,height=5cm]{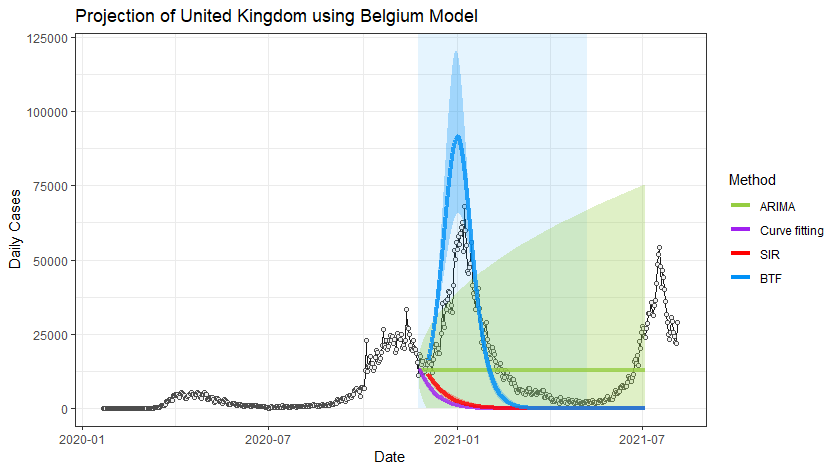}
\includegraphics[width=.45\textwidth,height=5cm]{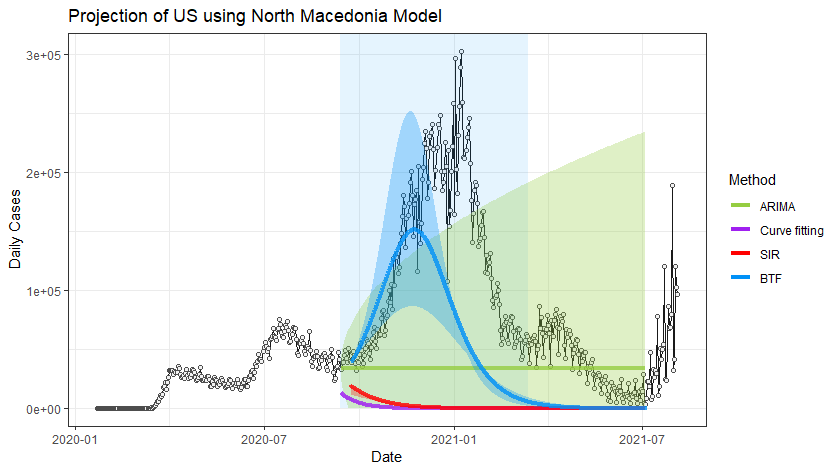}
\includegraphics[width=.25\textwidth,height=6cm]{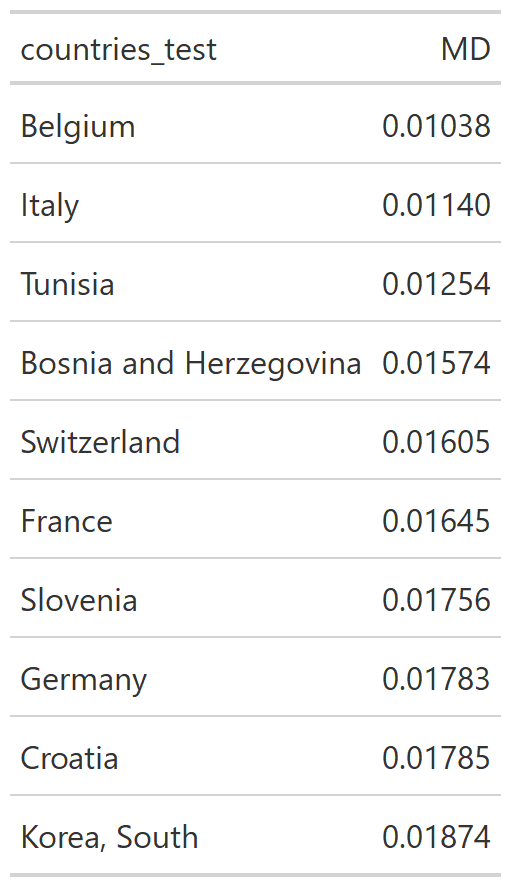}\hskip80pt
\includegraphics[width=.25\textwidth,height=6cm]{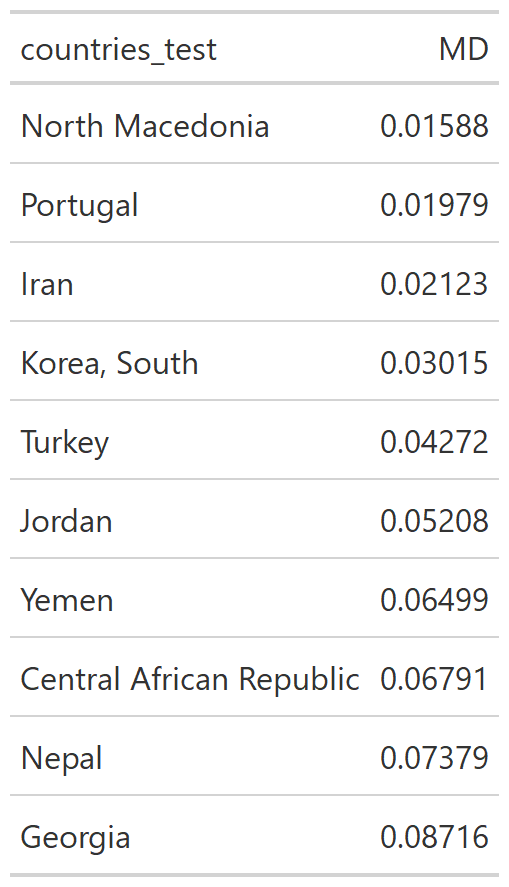}
\caption{Projection curves and matching country ranking tables for Singapore, South Africa, United Kingdom, and United States, using BFT projections.}
\label{group3}
\end{figure}

Figures \ref{group1}-\ref{group3} show four panels each with each panel depicting the following:  i) a observed daily infection curve; ii) BTF projected curves (solid blue curve) with sensitivity bands (darker blue shaded); iii) standard SIR projected curved (red) with sensitivity band (red shaded); iv) standard ARIMA($p,d,q$) forecast (green curve) with 95\% prediction interval (green shaded) and v) generalized logistic growth curve model with 95\% bootstrap prediction intervals (purple curve and shaded regions).  The time window of each surge of interest are the blue rectangular regions.   Underneath each country's plot is the matching table from which the coaching country's curve was derived.  

Making these kinds of projections is clearly a very challenging task and represent a type of aspirational goal (see \citet{Rosenfeld21}).  Thus, judging the accuracy of the BTF projections must be calibrated appropriately. For point estimate-based predictions, one can use absolute error; for interval-based predictions, the weighted integrated score is an option \citep{Rosenfeld21}. 

\begin{figure}[H]
\centering 
\includegraphics[width=.45\textwidth, height=7cm]{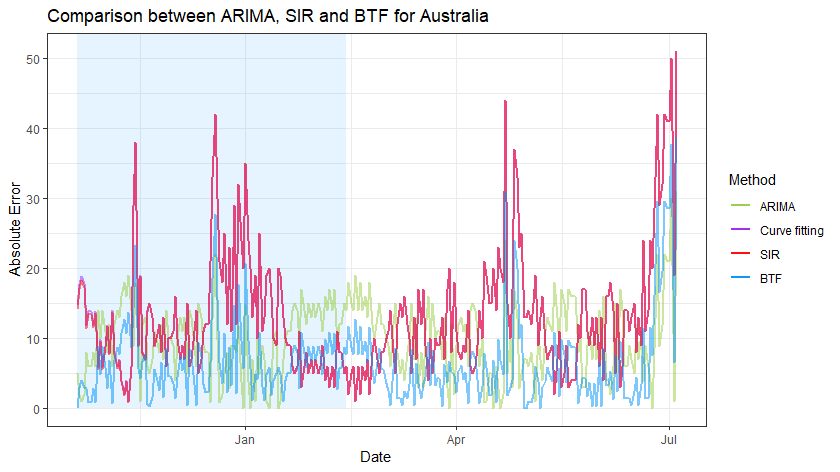}\hskip40pt
\includegraphics[width=.45\textwidth,height=7cm]{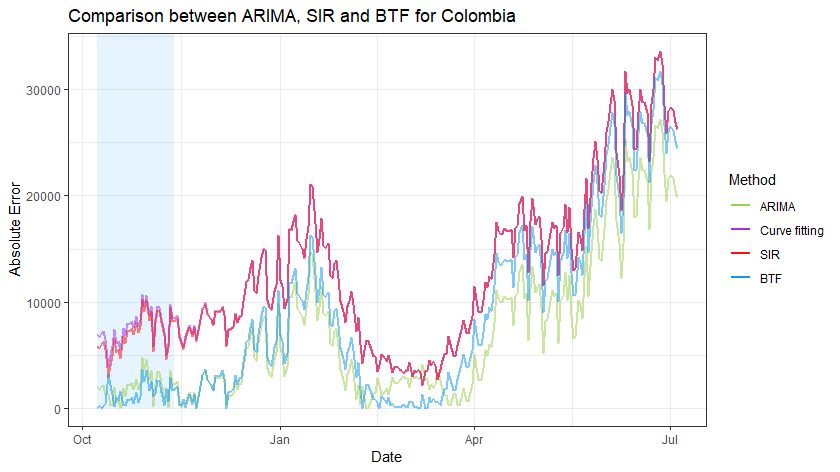}
\includegraphics[width=.45\textwidth,height=7cm]{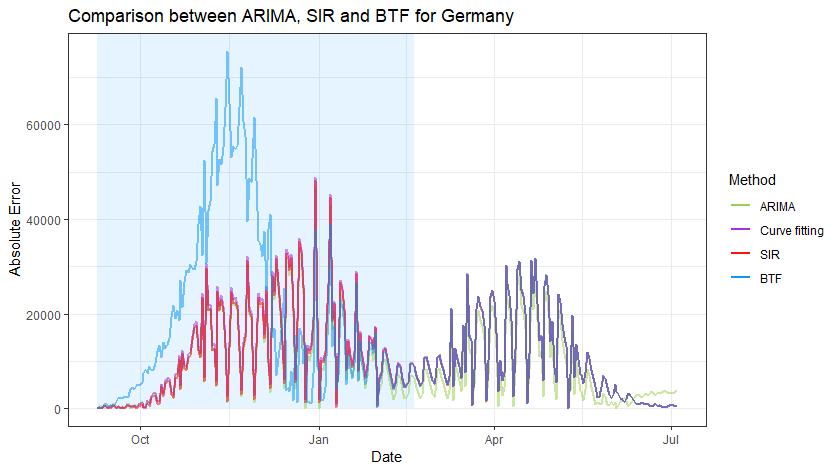}\hskip40pt
\includegraphics[width=.45\textwidth,height=7cm]{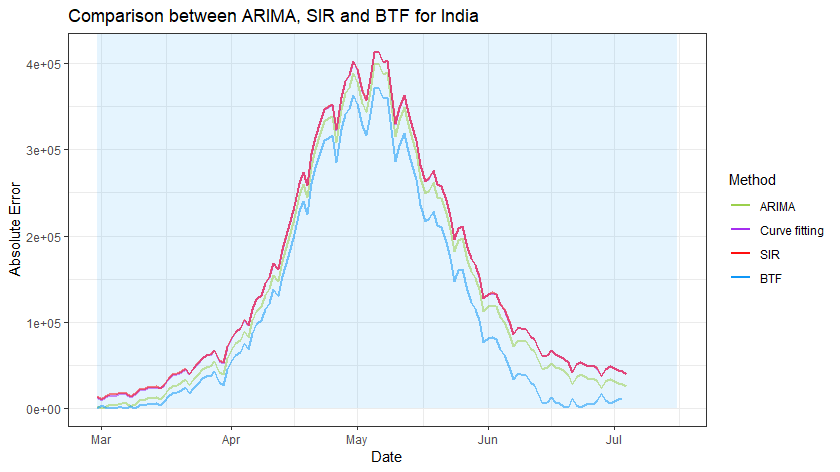}
\caption{AE estimates for Australia, Colombia, Germany, and India.}
\label{AEgroup1}
\end{figure}

\begin{figure}[H]
\centering 
\includegraphics[width=.45\textwidth, height=7cm]{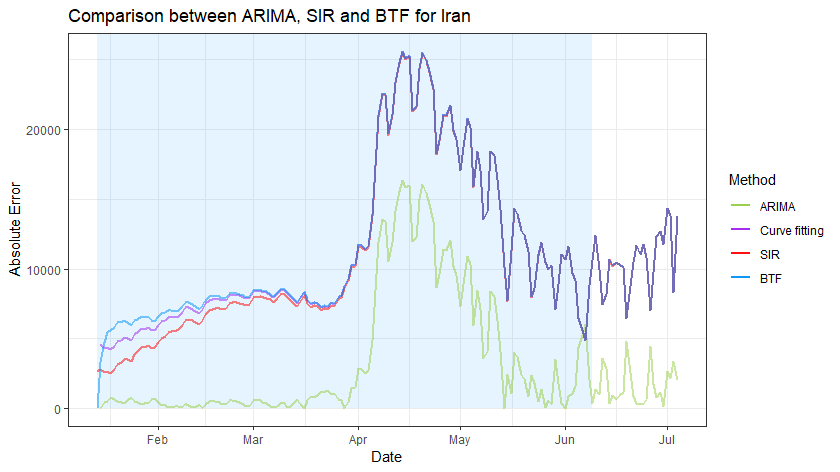}\hskip40pt
\includegraphics[width=.45\textwidth,height=7cm]{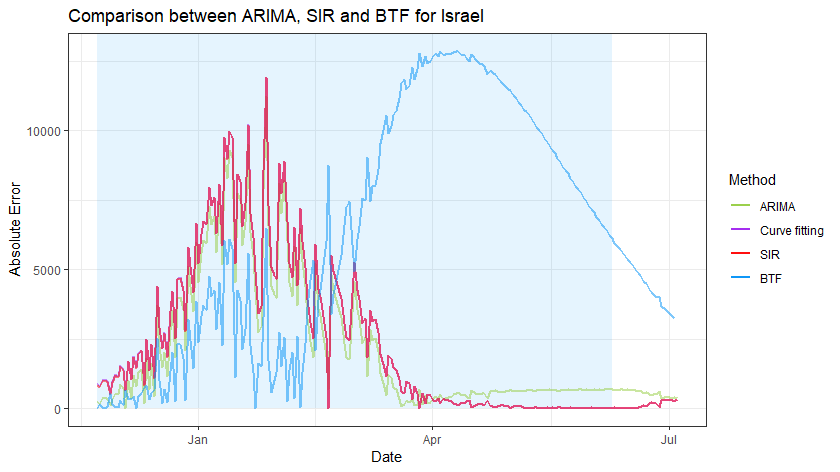}
\includegraphics[width=.45\textwidth,height=7cm]{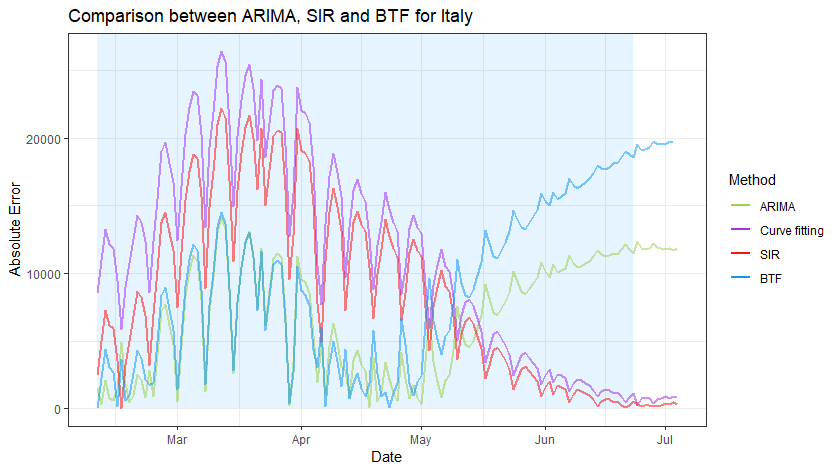}\hskip40pt
\includegraphics[width=.45\textwidth,height=7cm]{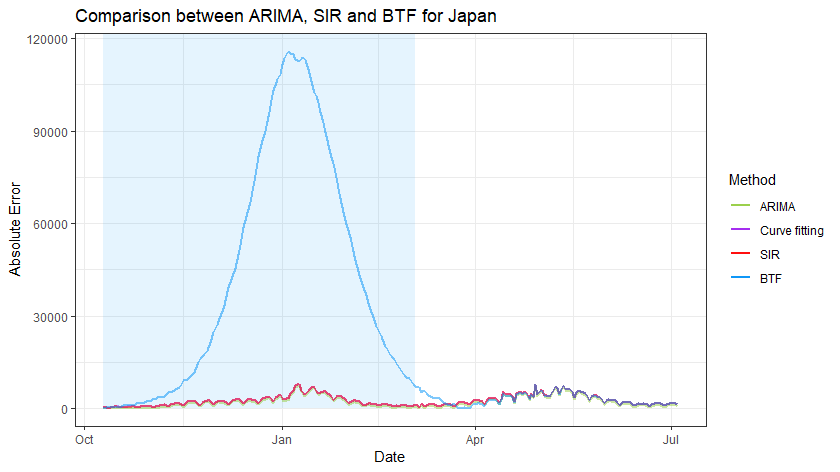}
\caption{AE estimates for Iran, Israel, Italy, and Japan.}
\label{AEgroup2}
\end{figure}

\begin{figure}[H]
\centering 
\includegraphics[width=.45\textwidth, height=7cm]{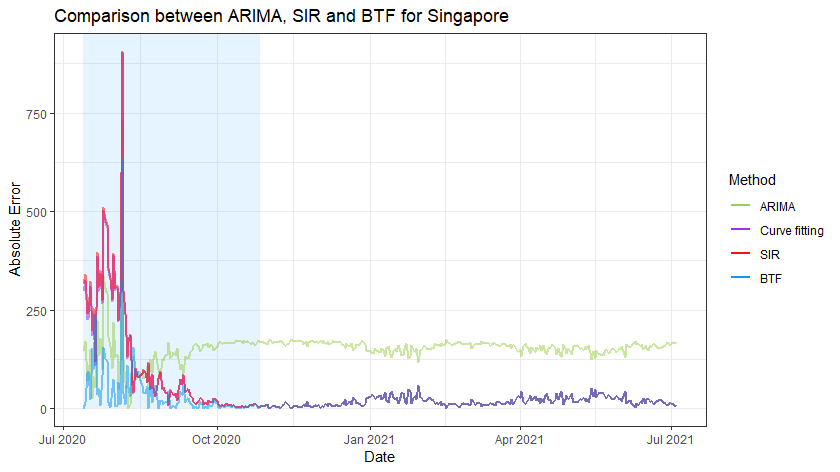}\hskip40pt
\includegraphics[width=.45\textwidth,height=7cm]{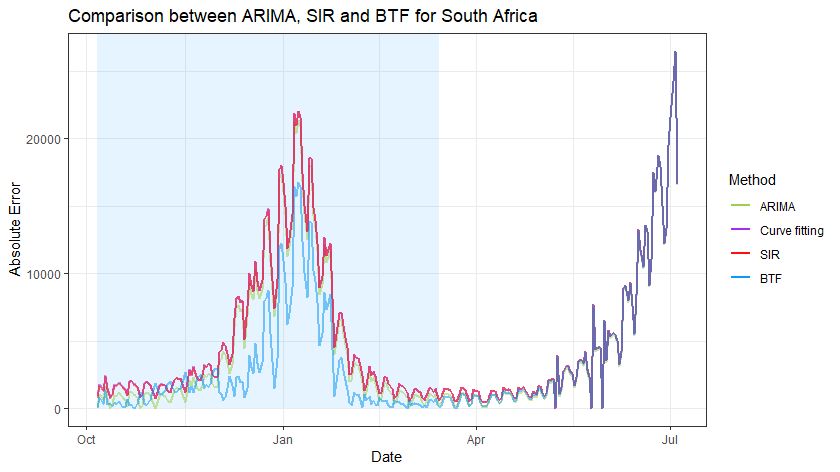}
\includegraphics[width=.45\textwidth,height=7cm]{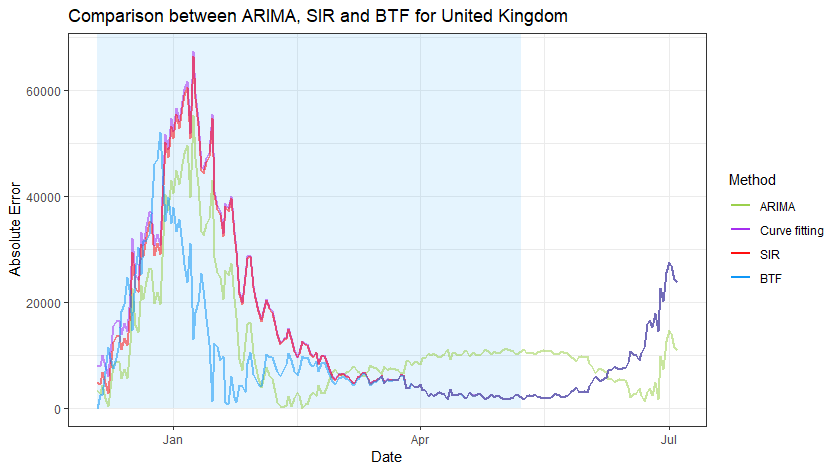}\hskip40pt
\includegraphics[width=.45\textwidth,height=7cm]{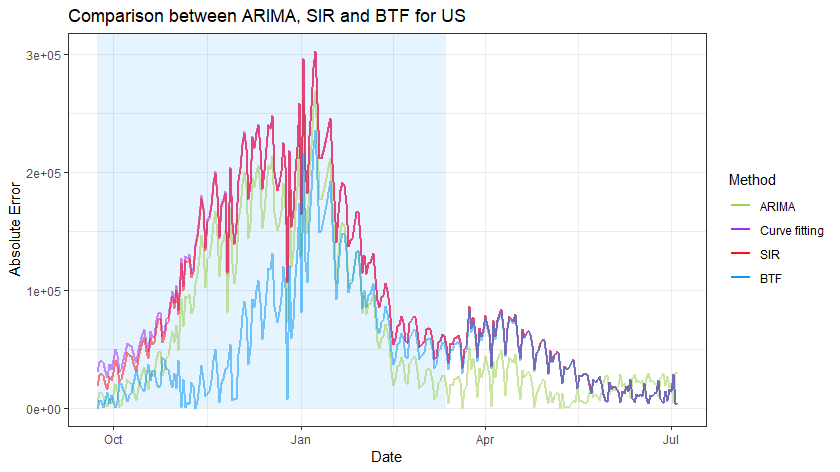}
\caption{AE estimates for Singapore, South Africa, United Kingdom, and the United States}
\label{AEgroup3}
\end{figure}

Figures~\ref{AEgroup1}-\ref{AEgroup3} show the absolute errors (AE) over time for BTF (blue curve) versus using 
the naive SIR model (red curve), ARIMA model (green curve) and generalized logistic growth curve (purple) projected forward from the same point in time. Once again, the light blue shaded rectangular regions correspond to the time windows of the future surge of interest.   Lower values of absolute error indicate a better fit to the actually observed future data. In $9/12$ circumstances the BTF's projections dominate naive model projections. The exceptions being Japan and Germany, where very large peaked surges were projected and much smaller peaked surge actually emerged; and Iran, where matched fits were poorly determined. For Israel, the BTF absolute error curve looks worse later in the shaded time window than earlier on; this corresponds to the projected peak for the surge being shifted too far to the right.  A similar ``flip'' in absolute error curves occurs for Italy.   For Australia, no differences were found but this country was the negative control. 

AE curves, while useful, do not convey other important information regarding surge projections.  For instance, it is of particular interest whether a surge projection accurately estimated peak height (within plus or minus 10 days) 
and/or location (within plus or minus 10 days).  Table \ref{matches} breaks this down for our analysis. It indicates that for 5 countries we did indeed achieve peak height match, and for 8 countries we achieved peak location match.  Contrast this to SIR and ARIMA models which worked only for Australia, the negative control.  

We also found an interesting result regarding India's projected surge.  The second surge corresponded to the emergence of the delta variant of COVID, which produced a peak height of over 400K daily infections.  Our projected estimate was only around 50K.  However, it has also been estimated that at surge peak, fully 90\% of the daily infection counts were attributable to the delta variant (\url{https://clingen.igib.res.in/covid19genomes/}). This means that 10\% came from other existing variants found in other countries.  Hence our projected peak approximates this number very accurately.  We actually would not expect to project a peak for a new variant accurately using BTF, since the method cannot accommodate new variants as currently formulated.

\begin{table}
\caption{Countries where BTF projections matched surge peak height and location. The * for India
under the peak height match column indicates the match is under a particular caveat described in the text.}
\label{matches}
\begin{tabular}{ll}
Peak Height Match & Peak Location Match \\ \hline
Israel            & Germany             \\
Australia         & Australia           \\
India$^*$            & India               \\
Singapore         & Singapore           \\
UK                & UK                  \\
                  & Japan               \\
                  & South Africa        \\
                  & US                 
\end{tabular}
\end{table}

\newpage
\newpage

\newpage

\clearpage

\newpage

\newpage

\clearpage

\section{Justification for the matching}

One of the features of a pandemic is that surges and recessions happen asynchronously across different countries.  We are relying on the fact that finding a best matched country by time-shifting to create overlayed earlier surges will in fact provide useful information to coach future surge projections of interest.  Thus it is necessary to say something regarding the optimality of this type of matching.

The correlation between asynchronous time series has been examined in what is termed lead-lag relationships between different financial markets \citep{deJong97}.  For example, a link has been established between index futures and the cash market where the futures market tends to lead the cash market (see for instance \citet{Stoll90}).  The analysis of information flows between markets on short varying time intervals is an active area of research.  In \citet{deJong97}, they developed a method for estimating correlations from irregularly spaced transactions data.  

For two stationary ARIMA processes, we can test for the presence of cross-correlation functions between the two asynchronous series.  However, this approach is sensitive to the choice of lag length and cannot tell the directionality of causality, only the presence or absence of it.  In addition, the statistic lacks power, as compared to regression-based tests discussed next.

One more clear way forward is to conduct a direct test for Granger causality \citep{Eichler13}, by regressing each variable on lagged values of itself and the other.  This can be written as
\begin{align*}
	Y_{t} = \eta_0 + \sum_{j=1}^{max(S_{1})-t} \eta_{j}Y_{t-j} + \sum_{k=1}^{max(S_{1,m})-t'} \kappa_{k}Z_{t'-k} + \epsilon_{t},
\end{align*}
where $\epsilon_{t}$ is white noise and $t'=t-l_{m}$, with $l_{m}$ the lag between $S_{1}$ and $S_{1,m}$.  Then Granger causality between the two asynchronous time series can be assessed by testing whether the $\kappa_{k}=0$ or not, using an F-test based on comparing nested residual sum of squares.

\subsection{Alternative strategy for matching using data enriched ARIMA models with lasso penalization}

Assume the daily infection counts for country $A$ in the currently ending surge are in time period $S_1$.  Assume a candidate country's $(B_m)$ daily infection counts during a previous surge earlier than country $A$'s currently ending surge are in time period $S_2$.  

Assume for $A$, 
that $Y_t$ follows the ARIMA($p,d,q$) model %
%
\eqref{ARIMA} with $t\in S_1$; and country $B_m$ also follows \eqref{ARIMA}, with $t\in S_2$,
but with autoregressive parameters $(\alpha_1 + \omega_{m,1}), \ldots, (\alpha_p +\omega_{m,p})$ and moving average parameters $(\theta_1+\nu_{m,1}), \ldots, (\theta_q+\nu_{m,q})$ (assuming $p$, $d$ and $q$ are the same for both).

Then the two can be pooled via shrinkage and weighting as in \citet{Chen15} as the solution to the penalized joint log likelihood, 
\begin{align*}
	l(A) + l(B_m) - \tau_{1} P(\omega_m) - \tau_{2} P(\nu_m),
\end{align*}
for penalty functions $P(\omega_m)$ and $P(\nu_m)$ and shrinkage parameters $\tau_1$ and $\tau_2$. Setting $P(\omega_m)= ||\omega_{m}||_{1}$ and $P(\nu_m)=||\nu_k||_1$ corresponds to joint lasso shrinkage \citep{Tibs96}.  The shrinkage parameters can be estimated as part of the penalized maximum likelihood estimation process.  

Then the best matching country would be,
\begin{eqnarray}
\tilde{B}_{m} = \text{arg\,min}_{k}(\lVert\hat{\omega}_{k}\rVert_{1} + \lVert\hat{\nu}_{k}\rVert_{1}).
\end{eqnarray}

\section{Connection to response coaching}
For country $A$, let's assume that $S_1$ defines the time period of the first surge, and $S_2$ the time period for the second surge.  Let $Y_t$ be the set of responses for country of interest $A$, $Z_{tm}$ be the corresponding set of responses for country $\tilde{B}_{m}$, and $S_{1,m}$ and $S_{2,m}$ define the time periods for $\tilde{B}_{m}$'s first and second surge. Note that $S_1 > S_{1,m}$ and $S_{2} > S_{2,m}$ by definition. 

Then let $f_{\delta_{2}}(Y_{t} \mid t \in S_2)$ be the fit for $A$ in $S_2$, indexed by parameter vector $\delta_2$; and $f_{\delta_{2,m}}(Z_{tm} \mid t \in S_{2,m})$ be the fit for $\tilde{B}_m$ in that country's $S_{2,m}$ indexed by parameter vector $\delta_{2,m}$.  Also let $h(t)$ be a function that maps from
$S_{2,m}$ to $S_2$.
These fits can be estimated using an SIR model.  Remember that the interval corresponding to $S_{2,m}$ is lagged with respect to the interval corresponding to $S_{2}$.  

Inspired by the response coaching idea of \citet{Tibs98}, we can write
\begin{align*}
	f_{\delta_{2,m}}(Y_{t},Z_{tm} | t \in S_{2}) = f_{\delta_{2,m}}(Y_{t} | t \in S_{2})f_{\delta_{2,m}}\left(Z_{tm} | h^{-1}(t) \in S_{2,m}\right),
\end{align*}
where $\delta_{2,m}$ is a coaching parameter vector specific to $S_{2,m}$ and shared with $A$ during its $S_{2}$.  Thus 
the prediction of $Y_t$ for $t \in S_2$ can be coached by country $\tilde{B}_m$ via the shared parameter vector $\delta_{2,m}$ and estimated by
\begin{align*}
	\hat{f}(Y_{t} | t \in S_{2}) = f_{\hat{\delta}_{2,m}}(Y_{t} | t \in S_{2}),
\end{align*}
where $\hat{\delta}_{2,m}$ is estimated from the fit of $\tilde{B}_{m}$ in $t \in S_{2,m}$ {\it and using the population characteristics of $A$ during $S_2$}.   

The fact that $S_1$ and $S_{1,m}$ are not the same, and that $S_2$ and $S_{2,m}$ are not the same, but that we are expecting $\tilde{B}_m$ to be informative regarding $S_2$ implies a periodic property of the pandemic across 
matched countries in time.  That is, $\tilde{B}_m$ represents a country that experienced a very similar first surge and thus there is information to be gleaned about predicting $Y$ by learning from $\tilde{B}_m$'s experience during
their (earlier) second surge.

\section{Discussion}
Making surge projections before the next surge begins is frankly necessary, given the highly infectious nature of many of the COVID-19 variants.  Waiting until after an inflection point will simply mean that one is always playing catch up against the virus.  

Our methodology attempts to do exactly this by employing a matching scheme to other candidate countries and then appealing to Granger causality in order to borrow from that matched country's observed ensuing daily case counts.  As we have discussed, once matching has occurred, the BTF projections themselves use a form of response coaching which can reduce variance over a non-coached model (Tibshirani and Hinton 1998).

It should be emphasized that our BTF projections cannot work well when a new variant emerges for the first time and is responsible as the driver of a new surge.  There is simply no hope to borrow strength from other countries.  This is the reason our projections for India were not accurate due to the first-time emergence of the COVID-19 delta variant in early 2021.  

So how can one know whether the BTF technology could be of use in a prospective sense?  One answer may lie in the recent work of \citet{Schioler21}, who developed a probabilistic model based on a hidden Markov model for infection spread and an approximation of a two stage sampling scheme to infer the probability of extinction of a current variant.  Should this probability be low, then BTF may be useful in projecting future surges.  Additional research is needed to adapt the methodology to allow for the possible emergence of new variants.

\section*{Reproducibility}
To ensure reproducibility, the codes used to generate BTF projections are available at \url{https://github.com/txl646/BTFcovid}.


\bibliographystyle{rss}
\bibliography{BTF}
\end{document}